\documentstyle{proceeding}
\input psfig.tex

\begin{document}
\title{The X-ray spectrum of the $\gamma$--bright quasar S5 0836+710}
\author{A. Comastri\inst{1,2} M. Cappi\inst{2} \and M. Matsuoka\inst{2}}  
\institute{Osservatorio Astronomico di Bologna 
 via Zamboni 33, I-40126 Bologna, Italy
\and The Institute of Physical and Chemical Research (RIKEN), 
2--1 Hirosawa, Wako, Saitama 351--01, Japan}
\maketitle
\begin{abstract}

We present the results of an ASCA observation of the high redshift
flat spectrum radio quasar S5 0836+710. 
The $\sim 0.4-10$ keV X-ray spectrum is remarkably flat with 
$\Gamma \sim 1.4$ and substantial intrinsic absorption at low energy.
The spectral slope was found to be in good agreement 
with the non--simultaneous ROSAT PSPC observations while no evidence
of intrinsic absorption has been found in the PSPC spectrum.
Our results suggest time variability of the absorbing material
on timescales less than five months in the source rest--frame.

\end{abstract}

\section{Introduction}

The flat spectrum radio quasar S5 0836+710 is a distant ($z$ = 2.172)
strong radio source ($S_{5 \rm GHz}$ = 2.67 Jy)
showing superluminal components in a bright, one-sided VLBI jet and high 
radio polarization (K\"uhr et al. 1981).
The optical polarization is low (1.1 $\pm$ 0.5\%; Impey \& Tapia 1990), while
strong optical variability has been detected in early 1992, 
including a rapid flare on February 16 (von Linde et al. 1993).
It is one of the most luminous sources 
discovered by CGRO--EGRET observatory in the hard 
$\gamma$--ray band, being the
highest redhift source 
among the about 50 blazars so far discovered by EGRET (von Montigny 
 et al. 1995).
The $\gamma$--ray spectrum in the energy range $\sim$ 30 MeV -- 2 GeV  
is very steep with a photon index
$\Gamma \simeq 2.4 \pm 0.2$ (Thompson et al. 1993).
Subsequent EGRET observations (Nolan et al. 1996), confirm 
the steep spectrum ($\Gamma = 2.62 \pm 0.36$) and significant flux 
 variability without associated spectral variations.

In the X-ray band S5 0836+710 has been previously observed by the ROSAT PSPC
in the $\sim 0.1-2.0$ keV band. 
The count rate decreased by a factor of $\sim$ 2 on a timescale of 
$\sim$ 7 months (observer frame) between the two pointed observations
(Brunner et al. 1994).
The soft X--ray spectrum is the flattest among the radio--loud quasars 
studied by Brunner et al. and is well
described by a single power law in both the observations 
with $\Gamma \simeq 1.40 \pm 0.05$.

The extreme X-- and $\gamma$--ray spectral properties of S5 0836+710
make this source an ideal target for the study of the high energy 
properties of blazars. In the following the results of the spectral analysis
of an ASCA observation are presented and compared with the ROSAT data.

\section{ASCA observation and data reduction}
  
ASCA observed S5 0836+710 on 1994 March 17 with the GIS for a 
total effective exposure time of about 16500 s and with the SIS for about 
10500 s. Due to an attitude problem at the beginnig of the observation the data
collected in the first 5 ksec have been removed from the present analysis.
The SIS was operated in 1 CCD mode and all the data were collected 
in FAINT mode and corrected for DFE and echo.
Source counts were extracted from a circle centered 
on the source of 6$^{\prime}$ radius for the GIS and
3$^{\prime}$ for the SIS.
Background counts were collected from the blansky files for GIS (from
a region uncontamined by NGC 6552 and at a similar off-axis distance of
the source) and from the same field of view for the SIS.
Backgrounds selected in a different way gave results consistent with
the analysis presented in the following.
Data preparation and analysis have been performed using the XSELECT package, 
and the extracted spectra were analyzed using version 8.5 of XSPEC.
 
\subsection{ASCA Spectral Analysis}

No significant variability has been detected in the GIS and SIS 
light curves, therefore the spectra have been collected for the whole
observation. The pulse height spectra were binned in order to get at least 
20 counts per bin for the GIS and SIS detectors and the adopted energy range 
were limited to 0.7--10 keV and 0.4--10 keV respectively.

A single absorbed power law was fitted to the data leaving first the column
density free to vary and then fixed at the Galactic value (Table 1).
We note that the two GIS (2 and 3) and the two SIS (0 and 1) give 
very consistent results; therefore we also fitted simultaneously both 
instruments. Confidence contours in the parameter space $N_H-\Gamma$, for
the simultaneous fit, are shown in Fig. 1. 

\begin{table}
      \caption{GIS and SIS - Power Law Fits}
         \label{KapSou}
      \[
           \begin{array}{cccc}
            \hline
            \noalign{\smallskip}
              Instrument &  \Gamma  &  N^{\rm a}_{\rm H} & \chi^{2} \cr
            \noalign{\smallskip}
            \hline
            \noalign{\smallskip}
            GIS2+3 & 1.42 \pm 0.10  &  < 14.7    & 0.89/241 \\
            GIS2+3 & 1.38 \pm 0.04  &  2.78^b    & 0.89/242 \\
            SIS0+1 & 1.46 \pm 0.08  & 12.2 \pm 3 & 0.79/257 \\ 
            SIS0+1 & 1.27 \pm 0.04  &  2.78^b    & 0.94/258 \\
            GIS+SIS & 1.45 \pm 0.05 & 11.4 \pm 3 & 0.87/503 \\ 
            GIS+SIS & 1.32 \pm 0.03 &  2.78^b    & 0.97/504 \\   
            \noalign{\smallskip}
             \hline
         \end{array}
      \]
\begin{list}{}{}
\item[$^{\rm a}$] Units of 10$^{20}$ cm$^{-2}$, $^{\rm b}$ Galactic column 
density.
\end{list}
   \end{table} 

The power law model provides a good fit to the GIS data leaving essentially
unconstrained the absorbing column density due to the limited low energy
GIS response ($>$ 0.7 keV). Significant evidence of intrinsic absorption was
found in the SIS data where a power law fit with the column density 
fixed at the Galactic value can be rejected at $>$ 99.9\% confidence
(F--test). Fitting the SIS data with $N_H \equiv N_{HGal}$, the slope becomes
flatter by $\Delta \Gamma \simeq 0.1$, and systematic residuals appear at soft
energies.
We have also performed spectral fitting taking into account the source redshift.
The best fit value for the absorbing column density in the source rest
frame is $N_H \simeq 1.2 \pm 0.4 \times 10^{22}$ cm$^{-2}$, while the power 
law slope remains flat, $\Gamma = 1.43 \pm 0.07$, up to $\sim$ 30 keV. 
An extrapolation at high energies of the
best fit power law requires a sharp spectral break of $\Delta \Gamma \sim 1$
in the range 4--8 MeV in order to be consistent with the EGRET measurements.
The derived luminosity in the 2--10 keV band quasar frame
is $\simeq$ 6 $\times$ 10$^{47}$ ergs s$^{-1}$ corresponding to an 
observed flux of $\simeq$ 1.4 $\times$ 10$^{-11}$ ergs cm$^{-2}$ s$^{-1}$.
No evidence of an iron emission line is present in the GIS or SIS 
spectra with a tight upper limit for the equivalent width 
EW $\leq$ 31 eV (90\% confidence). 

\subsection{ASCA and ROSAT Spectral Analysis}

To better compare the ASCA spectral fitting results with the 
ROSAT PSPC data we have extracted
from the public archive the two available pointed observations 
(March 92 and November 92).
The two ROSAT observations give similar spectral parameters allowing 
a simultaneous spectral fitting. 
The best fit value for the absorbing column density is consistent 
with the Galactic one (Fig. 1) and the derived power law slope 
$\Gamma = 1.41 \pm 0.04$ with $N_H \equiv N_{HGal}$ is 
in very good agreement with the findings of Brunner et al. (1994).
>From a comparison of the
confidence contours in the $N_H-\Gamma$ parameter space, 
we find that the power law slopes agree
within 90\%, but the column densities obtained with ROSAT and ASCA are
significantly different. This large difference, a factor of $\sim$ 4,
cannot be accounted for in terms of a systematic error in 
the SIS low energy calibration estimated to be 
$\sim 3 \times 10^{20}$ cm$^{-2}$.

\begin{figure}
\psfig{figure=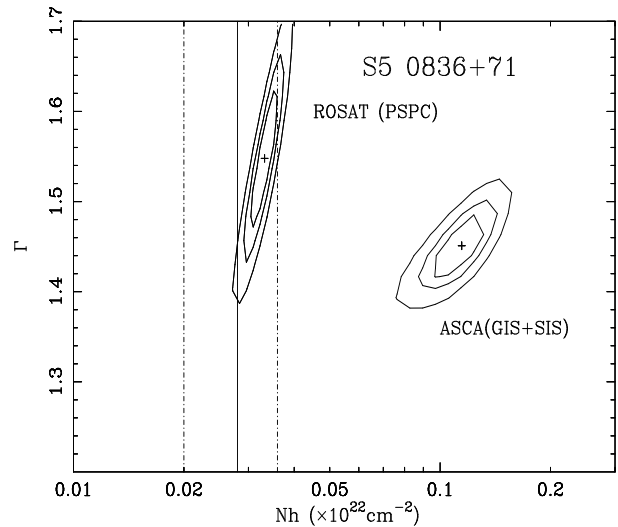,height=7.5truecm,width=9truecm,angle=-90}
\caption[]{68\%, 90\%, 99\% confidence contours in the $N_H-\Gamma$ plane. 
The vertical lines indicate the Galactic column density and estimated 30\% error
(Dickey \& Lockman 1990).}
\end{figure}

\section{Discussion} 

Even assuming the ASCA SIS systematic uncertainty in determining 
the soft X-ray absorption column, 
the spectrum of S5 0836+710 requires absorption in excess to the Galactic value.
Alternative models have been considered as well. A broken power law cannot
be excluded with the present data but seems unlikely because the strong 
depletion of the data in the lower channels requires an unplausibly 
reversed slope with $\Gamma < 0.4$ in the soft energy band. 
A full discussion of this and other models is
deferred to Cappi et al., in preparation. 
The ASCA spectrum, combined with the ROSAT data, provide a strong 
evidence for variable absorption in the direction of S5 0836+710
on a time scale $<$ 5 months (quasar frame).
S5 0836+710 would not be the first high redshift radio-loud
quasar which exhibit intrinsic absorption (Elvis et al. 1994)
or variable absorption (NRAO 140, Marscher 1988). 
A Galactic origin for the variable absorption in NRAO 140 
has been suggested by Turner et al. (1995) combining the
atomic and molecular (CO) column densities 
along the line of sight to the quasar, through the 
Perseus cloud complex.
The high Galactic latitude and the lack of CO emission 
in the direction of S5 0836+710 (Liszt et al. 1993)
strongly support an intrinsic origin for the variable absorption.

\vskip 0.4cm
\begin{acknowledgements}
AC acknowledges hospitality at the RIKEN Cosmic Radiation Laboratory
where part of this work has been done and Dr. T. Di Girolamo for useful
discussions.
\end{acknowledgements}

\end{document}